\documentclass[sigconf]{acmart}

\newcommand{\tcmd}[2]{\textbf{\texttt{#1}}{\textit{#2}}}
\newcommand{\eg}[0]{\textit{e.g.,}}
\newcommand{\ie}[0]{\textit{i.e.,}}

\AtBeginDocument{%
  }

\copyrightyear{2023}
\acmYear{2023}
\setcopyright{licensedusgovmixed}\acmConference[HRI '23 Companion]{Companion of the 2023 ACM/IEEE International Conference on Human-Robot Interaction}{March 13--16, 2023}{Stockholm, Sweden}
\acmBooktitle{Companion of the 2023 ACM/IEEE International Conference on Human-Robot Interaction (HRI '23 Companion), March 13--16, 2023, Stockholm, Sweden}
\acmPrice{15.00}
\acmDOI{10.1145/3568294.3580112}
\acmISBN{978-1-4503-9970-8/23/03}

\usepackage{stfloats,microtype,balance,enumitem}
\usepackage[some,bottom]{background}

\begin{document}

\title{Crowdsourcing Task Traces for Service Robotics}


\author{David Porfirio}
\orcid{0000-0001-5383-3266}
\affiliation{%
  \institution{NRC Postdoctoral Research Associate}
  \institution{Naval Research Laboratory}
  \city{Washington}
  \state{DC}
  \country{United States}
  \postcode{53706}
}
\email{david.porfirio.ctr@nrl.navy.mil}

\author{Allison Sauppé}
\orcid{0000-0002-7548-368X}
\affiliation{%
  \institution{University of Wisconsin–La Crosse}
  \city{La Crosse}
  \state{Wisconsin}
  \country{United States}
  \postcode{54601}
}
\email{asauppe@uwlax.edu}

\author{Maya Cakmak}
\orcid{0000-0001-8457-6610}
\affiliation{%
  \institution{University of Washington}
  \city{Seattle}
  \state{Washington}
  \country{United States}
  \postcode{98195}
}
\email{mcakmak@cs.washington.edu}

\author{Aws Albarghouthi}
\orcid{0000-0003-4577-175X}
\affiliation{%
  \institution{University of Wisconsin–Madison}
  \city{Madison}
  \state{Wisconsin}
  \country{United States}
  \postcode{53706}
}
\email{aws@cs.wisc.edu}

\author{Bilge Mutlu}
\orcid{0000-0002-9456-1495}
\affiliation{%
  \institution{University of Wisconsin–Madison}
  \city{Madison}
  \state{Wisconsin}
  \country{United States}
  \postcode{53706}
}
\email{bilge@cs.wisc.edu}

\renewcommand{\shortauthors}{David Porfrio, Allison Sauppé, Maya Cakmak, Aws Albarghouthi, \& Bilge Mutlu}
\renewenvironment{quotation}
{\list{}{\leftmargin=10pt
  \listparindent \parindent
  \itemindent \listparindent
  \rightmargin \leftmargin
  \parsep \parskip}%
  \item\relax\noindent\ignorespaces}
{\endlist}
\SetBgContents{Distribution Statement A: Approved for public release. Distribution unlimited.}
\SetBgScale{1}
\SetBgOpacity{1}
\SetBgVshift{1cm}
\SetBgColor{black}

\begin{abstract}
Demonstration is an effective end-user development paradigm for teaching robots how to perform new tasks. In this paper, we posit that demonstration is useful not only as a teaching tool, but also as a way to understand and assist end-user developers in thinking about a task at hand. As a first step toward gaining this understanding, we constructed a lightweight web interface to crowdsource step-by-step instructions of common household tasks, leveraging the imaginations and past experiences of potential end-user developers. As evidence of the utility of our interface, we deployed the interface on Amazon Mechanical Turk and collected 207 task traces that span 18 different task categories. We describe our vision for how these task traces can be operationalized as task models within end-user development tools and provide a roadmap for future work.
\end{abstract}



\maketitle

\section{Introduction}
\begin{figure}[!t]
    \centering
    \includegraphics[width=\columnwidth]{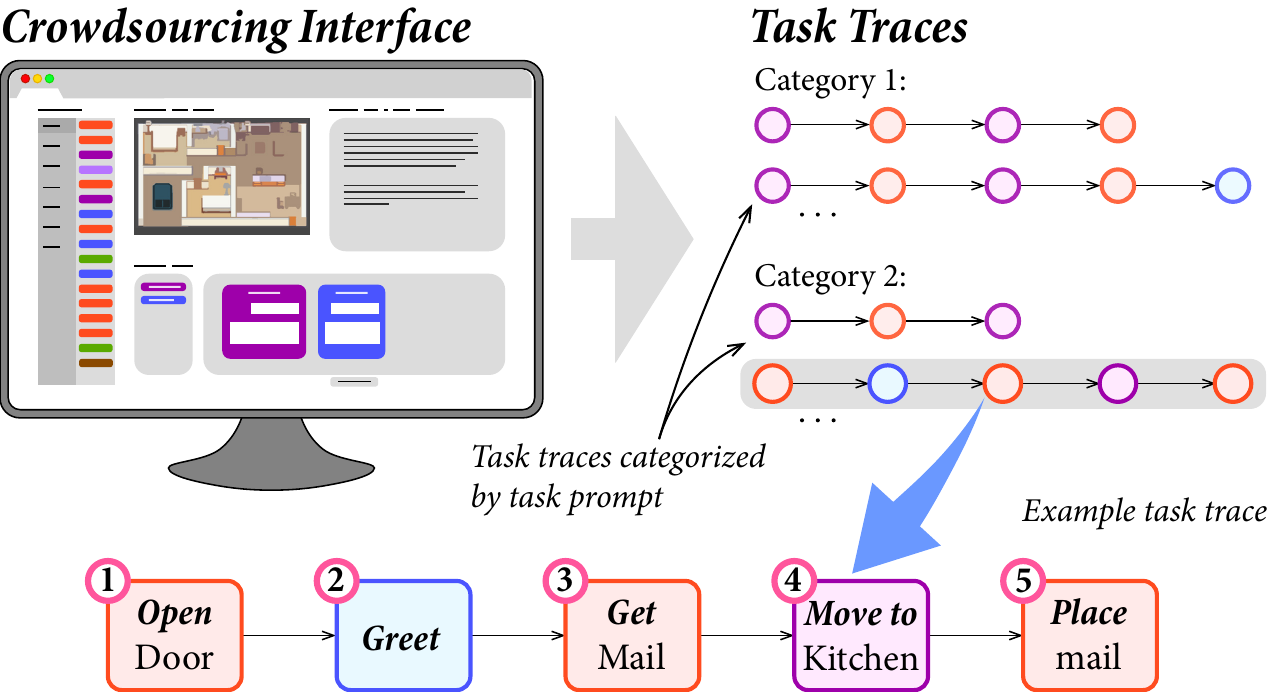}
    \caption{Our crowdsourcing interface collects individual task traces, which are organized by task category.}
    \label{fig:teaser}
\end{figure}

End-user developers who script personalized service robot applications face numerous challenges related to the unrestricted environments these robots often traverse, the social nuances that many of these robots must navigate, and a lack of technical expertise or appropriate development systems and tools required to address these complexities. To address these challenges, we envision end-user developer tools (such as \cite{glas2016ic, cao2019vra, huang2017code3}) using built-in task models that capture the high-level flow of common household tasks to transfer task knowledge to end-user developers. To illustrate our vision, consider an end-user developer who wishes to create a reusable, personalized task script to retrieve mail from the mailbox. Using a spoken-language interface as an example, if the end user specifies ``Get the mail,'' the interface should leverage a \textit{fetch} model to propose a plausible next step in the task---bring the mail to a convenient location for the end user to access later.

As a first step toward realizing our vision of transferring task knowledge to end-user developers,
we (1) constructed a web interface for crowdsourcing demonstrations, or \textit{traces}, of the step-by-step flow of common service tasks and (2) crowdsourced a preliminary dataset of 207 traces grouped within 18 separate task categories. In designing our crowdsourcing approach, our requirements were threefold. First, traces should not be tied to any particular context. As such, our trace collection interface, shown in Figure \ref{fig:web_app}, presents crowdworkers with task prompts that contain minimal contextual details and encourages crowdworkers to rely heavily on their own imaginations and past experiences. Second, the collection of traces should be efficient and scalable to any imaginable service task in the home or workplace. Therefore, crowdworkers using our collection interface need only to designate \textit{what} critical steps in a task are performed rather than \textit{how} they are performed. Finally, traces should capture the different ways that end-user developers might personalize tasks rather than the ground truth for how a robot \textit{should} perform these tasks.

Our contributions are shown in Figure \ref{fig:teaser} and include (1) the design of an interface to collect discrete, decontextualized, and personalized task traces and 
(2) a dataset that we collected by deploying this interface on Amazon Mechanical Turk (MTurk).\footnote{\url{https://www.mturk.com/}} 
\section{Related Work}

Our work draws from existing interfaces for simulating and demonstrating robot tasks and prior work with datasets and models that capture human activity and human-robot interaction. 

\subsection{Robot Simulation \& Demonstration Tools}\label{sec:simulators}

Many existing data collection and simulation interfaces for robotics have realistic environments and physics engines, such as \textit{Habitat 2.0} \cite{szot2021habitat} and \textit{iGibson 2.0} \cite{li2021igibson}. \textit{iGibson 2.0} additionally maps simulations to discretized, logical states that can be useful for programming or planning tasks. The \textit{SEAN 2.0} simulator represents a further step in simulation in its ability to model the behavior of pedestrians in a scene \cite{tsoi2022sean}.
Social behaviors are similarly present within the online game interface proposed by \citet{chernova2011crowdsourcing}, in which 
 similar to our collection interface, interaction data is crowdsourced.
An additional close match to our own needs is the \textit{VirtualHome} simulator in its ability to crowdsource discretized, step-by-step demonstrations of tasks and social activities \cite{Puig_2018_CVPR}. However, we require crowdworkers to rely on their own past experiences or imaginations rather than being provided with a realistic simulated context, thus ruling out many modern simulators such as \textit{VirtualHome}.

\subsection{Task Datasets \& Models}
Prior work has produced datasets and models that capture demonstrations of common household and workplace tasks, in addition to human-human and human-robot interactions. While much previous work has focused on collecting data to characterize rich multimodal sensing and behaviors \cite[\eg{}][]{yasser2008, ben2017ue, dinesh2013}, we focus instead on data that captures the discrete ordering of events in a task or interaction.

Often, such data arises from studying human behavior in the laboratory. From observing eight interaction dyads participate in five different scenarios---conversation, collaboration, instruction, interviewing, and storytelling---\citet{sauppepaper2014} extracted models of interaction patterns in a vein similar to \citet{kahn2008patterns}. Using the same observations, \citet{sauppePhDdissertation} describes larger-scale models that capture the overall flow of each scenario. These laboratory-generated datasets encompass a small set of interaction scenarios, however, while we require data spanning a wider range of tasks.

Datasets also arise from outside of the laboratory, such as the \textit{Loqui} dataset of transcribed human-human conversational interactions \cite{loqui}. Also obtained in the wild, the \textit{ARAS} \cite{hande2013} and \textit{Orange4Home} datasets \cite{cumin2017dataset} characterize daily human activity using passive sensor data. While there is immense potential for human task and activity models to arise from this prior work, these datasets are limited in scope or capture human activity at too low of detail for characterizing individual tasks. Recent advances in large-scale in-the-wild datasets, in contrast, have proven effective in training robots to perform novel tasks \cite{brohan2022rt}. However, our goal is not to transfer task skills directly to a robot; rather, we wish to capture how end-user developers imagine themselves specifying a task.

Various datasets have also been collected through simulation, such as the \textit{ALFRED} dataset that consists of automatically-generated task demonstrations \cite{Shridhar_2020_CVPR}. \textit{VirtualHome} has also been used to collect demonstrations of a wide range of household tasks and social interactions \cite{Puig_2018_CVPR}, which in contrast to \textit{ALFRED}, are human-generated. Existing datasets generated in simulation, however, suffer from the same drawbacks discussed in \S\ref{sec:simulators}, namely being influenced by contextual characteristics enforced by the simulator.

\section{Trace Collection}\label{sec:proc}

In this section, we describe our crowdsourcing interface, our procedure for collecting task traces, and the results of data collection.\footnote{ Portions of \S\ref{sec:proc} were presented in Chapter 7 of the first author's Ph.D. thesis \cite{porfirio2022authoring}.}\textsuperscript{,}\footnote{Our study materials, code for the web interface, and resulting dataset can be found at \url{https://osf.io/jt9hr}}

\subsection{Collection Interface}
\begin{figure*}[!t]
  \centering
  \includegraphics[width=\textwidth]{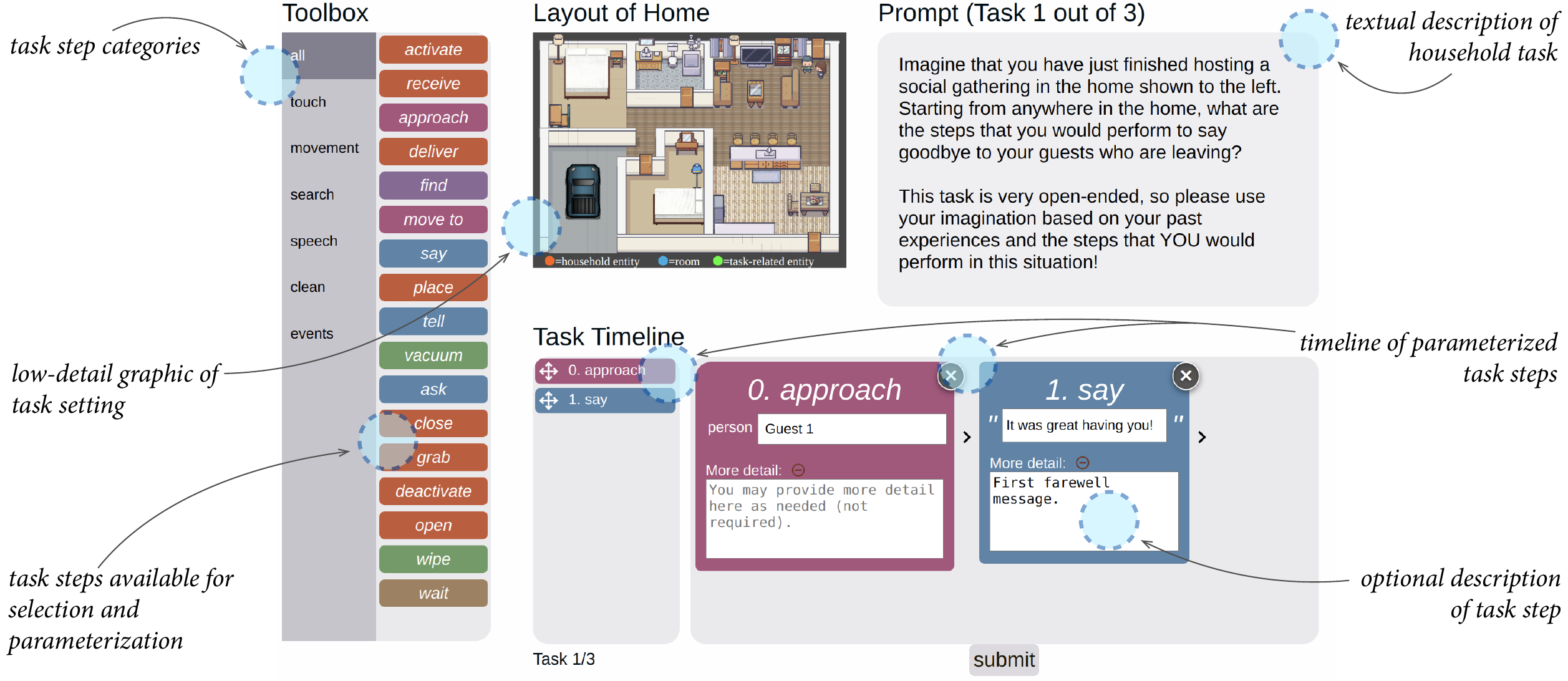}\vspace{6pt}
  \caption{The web interface that we created to collect task traces. The Layout of Home pane uses graphics from LimeZu.\protect\footnotemark}\vspace{-6pt}
  \Description{The web interface for collecting templates on Amazon Mechanical Turk. The Layout of Home pane uses graphics from LimeZu.}
  \label{fig:web_app}
\end{figure*}

Figure \ref{fig:web_app} depicts the crowdsourcing interface that we deployed on MTurk to collect task traces. Within the interface, the components ``Prompt'' (Figure \ref{fig:web_app}, top right) and ``Layout of Home'' (Figure \ref{fig:web_app}, top center) describe a category of household tasks or social scenarios that crowdworkers should imagine themselves completing. The prompt provides a textual description of the scenario, while the layout is intended to assist crowdworkers in understanding and situating themselves within the prompt. The layout is minimally interactive, allowing crowdworkers to hover over it and receive information about potentially relevant rooms or entities within the home. The prompt and layout are purposefully low in detail in order to stimulate imagination and the recollection of past experiences to fill in missing task details. \S\ref{subsec_procedure} provides a list of the 18 prompt categories we used within our trace collection procedure.

Crowdworkers respond to a prompt by dragging task \textit{steps} from the ``Toolbox'' component and dropping them into the ``Task Timeline'' to create a task trace. Table \ref{table:steps} defines the 17 parameterizable steps available in the interface (Figure \ref{fig:web_app}, left), which are intended to map to robot skills. The interface provides descriptions of each step as tooltips.

\begin{table}[b]
\begin{center}
\begin{tabular}{ l c l }
 \hline
 \tcmd{move to:}{target} & --- & move to a \textit{target} \\
 \tcmd{find:}{target} & --- & search for a \textit{target} \\
 \tcmd{grab:}{item} & --- & grab an \textit{item} \\
 \tcmd{open:}{container} & --- & open a \textit{container} \\
 \tcmd{close:}{container} & --- & close a \textit{container} \\
 \tcmd{deliver:}{item, target} & --- & bring an \textit{item} to a \textit{target} \\
 \tcmd{receive:}{item} & --- & receive an \textit{item} from someone \\
 \tcmd{place:}{item, container} & --- & place an \textit{item} in a \textit{container} \\
 \tcmd{approach:}{person} & --- & approach a \textit{person} \\
 \tcmd{say:}{exact-speech} & --- & say the \textit{exact speech} as specified \\
 \tcmd{tell:}{story} & --- & tell a \textit{story} \\
 \tcmd{ask:}{exact-speech} & --- & ask a question using \textit{exact speech} \\
 \tcmd{activate:}{device} & --- & turn a \textit{device} on \\
 \tcmd{deactivate:}{device} & --- & turn a \textit{device} off \\
 \tcmd{vacuum:}{room} & --- & clean a \textit{room} by vacuuming it \\
 \tcmd{wipe:}{surface} & --- & clean a \textit{surface} by wiping it \\
 \tcmd{wait}{} & --- & wait for something to happen \\
 \hline
\end{tabular}
\end{center}
\caption{The parameterizable steps available within our crowdsourcing interface.}
\label{table:steps}
\end{table}

To instantiate a step, crowdworkers click on it or drag it from the toolbox to the timeline, after which it becomes available to be parameterized. Figure \ref{fig:web_app} (bottom) displays an instantiated \tcmd{approach}{} step, in which the \textit{person} parameter is parameterized with the \textit{Guest 1} argument. Therefore, the first step in the trace is \tcmd{approach:}{Guest 1}. To facilitate open-endedness in addressing the prompts, parameterization is free response. We provided an additional free response text box at the bottom of each instantiated step to allow crowdworkers to provide additional detail or justify a particular step. 

\subsection{Collection Procedure}\label{subsec_procedure}
We conducted an IRB-approved study in which 105 MTurk crowdworkers (\textit{Turkers}) used our collection interface. To participate, Turkers needed IP addresses geographically within the United States and a >95\% task approval rate. In providing consent to participate, Turkers were informed that their data would be publicly shared and that their responses were subject to approval by the research team. Turkers whose work we approved were paid \$2.67 for an expected completion time of 20 minutes.

After giving their consent to participate, Turkers were directed to a tutorial web page that described via text and a three-minute YouTube video tutorial how to use the interface and the criteria for their responses to be approved. Our approval criteria were such that (1) Turkers could not provide traces with only one step, and (2) traces must address the provided task prompts. We encouraged Turkers to use the optional free response description text boxes under each step in the timeline to justify their work (\ie{} if a response outwardly seemed irrelevant to the prompt). Generally, we did not discard traces from Turkers who provided justifications. If fewer than two traces provided by a Turker failed to meet the acceptance criteria outlined on the tutorial page, we discarded all of the Turker's traces. 
\footnotetext{\url{https://limezu.itch.io/}}

We used the YouTube IFrame Player API\footnote{\url{https://developers.google.com/youtube/iframe_api_reference}} to ensure that Turkers watched the video before proceeding to the collection interface. Once allowed to proceed, each was asked to respond to three separate task prompts. An example \textit{Mail} prompt is as follows:

\begin{quotation}
  \textit{Imagine that you live in the home shown to the left, and the mail has just arrived through a slot in the front door. Before opening any of the letters or packages and starting from anywhere in the home, what steps would you take to fetch the mail?}
\end{quotation}

Each prompt then ends with the statement \textit{``This task is very open-ended, so please use your imagination based on your past experiences and the steps that YOU would perform in this situation!''}

In addition to the \textit{Mail} prompt, an additional 17 prompts captured tasks within the following categories: \textit{Greeting}, \textit{Farewell}, \textit{Groceries}, \textit{Storytelling}, \textit{Alarm}, \textit{Announcement}, \textit{Vacuum}, \textit{Answer Door}, \textit{Turn on Lights}, \textit{Delivery}, \textit{Ask About Day}, \textit{Phone Call}, \textit{Patrol}, \textit{Find}, \textit{Dust}, \textit{Declutter}, and \textit{Answer Question}. In designing each prompt, we aimed to include a spectrum of both social and non-social tasks and intended for the prompts to be general enough to allow crowdworkers immense freedom in how they chose to respond.

\subsection{Collection Results}
Of the first 69 Turkers, we rejected 12 ($82.6\%$ approval rate). Of the remaining 36, who participated at later dates than the initial 69, we noticed a substantial increase in spam responses and only approved 13 ($36.1\%$ approval rate). We observed no difference in the quality of approved responses after the increase in spam. Of the Turkers whose work we include in the dataset, three provided a single trace that did not meet our approval criteria, so we discarded these individual traces and kept the remainder of their work.

The final dataset includes 207 traces sourced from 70 Turkers. On average, the number of traces collected per task category is 11.5 ($\mathrm{min}=10$ traces, $\mathrm{max}=16$ traces). On average, traces contained  6.23 steps ($\mathrm{min}=2$ steps, $\mathrm{max}=23$ steps). Turkers appeared attentive and thoughtful when using the interface---62 Turkers used the free response boxes on their instantiated steps to provide additional detail, with the total number of descriptions being $706$ (a rate of $0.55$ descriptions per step, with $1289$ total steps provided). 

Participants were additionally asked for feedback on the interface after they finished responding to all three prompts, and their positive feedback further indicates high engagement. Examples of feedback include the following:
\begin{quotation}
\textit{This was one of the more unique tasks I've done on mturk. It was very interesting.}
\end{quotation}

\begin{quotation}
\textit{I would love to do more of these! It reminds me of solving problems using pseudocode!}
\end{quotation}

\begin{quotation}
\textit{I enjoyed it, it felt like a roleplaying game.}
\end{quotation}

The feedback also indicates that some Turkers were uncertain as to whether they completed the task correctly:

\begin{quotation}
\textit{I believe I did well on this task but would appreciate any feedback.}
\end{quotation}

\begin{quotation}
\textit{I would have liked to receive feedback during the exercise to know if I was hitting the mark and accomplishing the goal of the study. I tried my best.}
\end{quotation}

Other interesting phenomena occurred during the collection of traces. $61.4\%$ of Turkers used the \tcmd{wait}{} step, which was often accompanied by free response descriptions of external events that must occur in order for the task to proceed (\eg{} waiting for a verbal response). Additionally, many traces explicitly resolve preconditions that an autonomous robot could likely resolve by itself. In the step \tcmd{wipe: }{table}, for example, a robot is likely aware of the necessary precondition of being in possession of a duster or cloth, and if it does not have one, of the need to fetch one. Turkers were not aware that their traces were intended to be used for service robots and often explicitly included these preconditions in their traces.
\section{Discussion}

In this work, we provide a lightweight approach for crowdsourcing task traces with the goal of transferring knowledge from the traces to end-user developers who script tasks for service robots. 
As developers specify the high-level steps for a robot to perform (\eg{} a command to put a bag of groceries in the kitchen) a model of the task at hand can suggest ways that the task specification can be further personalized, such as by inferring edits to the task flow or by prompting clarification on unstated task structure. For example, a developer tool may suggest a \texttt{foreach} loop to an end user specifying a \textit{Groceries} task.

In transferring task knowledge to end-user developers, we are inspired by previous work in which program hints provided by a developer serve as input to an automated program synthesizer for completion \cite[\eg{}][]{srivastava2013template,solar2008program,solar2009sketching,solar2013program}. In our case, we envision a human-in-the-loop pipeline in which end-user developers provide a set of task steps as hints, and a developer tool uses a task model constructed from task traces to suggest where additional steps, loops, or branch points might be needed. This vision is closely aligned with prior work in human-robot interaction, in which \textit{templates} have been used as pre-existing generic, reusable program specifications to be selected and instantiated by end users \cite{ferrarelli2017design}.

To achieve our vision, we aim to create task models from individual traces (such as in \cite{mohseni2015interactive}) or multiple task traces. A na\"{i}ve approach could treat individual traces as models themselves and compute a ``diff'' between the hint and the trace to find missing steps omitted by the end user. In a more sophisticated approach, multiple traces under the same category could be combined into a probabilistic model, such as a Markov chain, that could be used to compute the probability of a particular step, loop, or branch point being present in the task. If there is uncertainty in the task hint, such as if the end user provides an ambiguous or incomplete step (\eg{} the spoken language utterance ``that goes over there''), combined traces may also serve as hidden Markov models, in which the current step of the task being specified by the end user must be inferred.

Our dataset currently limits our ability to pursue the more sophisticated, probabilistic approach, as we would likely need to collect many more traces and rigorously post-process these traces to remove noise. 
Our dataset is further limited in the number of task categories, currently 18, within which we sourced traces.
Future work must therefore involve (1) collecting more task traces per task category in order to sufficiently build models that represent the different ways that the same type of task can be performed, and (2) modifying our interface to streamline the collection of more traces.

Although our interface is lightweight and can scale to new task categories, it too has limitations that hinder data collection. First, in order to introduce a new task category, a researcher must manually construct a new prompt, potentially injecting their own biases. Future work should enable task categories to naturally emerge from the collection of crowdworkers' daily routines. Additionally, at present time, individual traces provided by crowdworkers must be manually screened by a researcher. Although we believe that our trace acceptance criteria are highly objective, future work must strive toward more systematic criteria, preferably automated so as to remove any possible researcher bias or error. 

\section{Conclusion}
We present a web interface for collecting traces of service robot tasks and a small dataset collected from the deployment of this interface on Amazon Mechanical Turk. We describe our vision of aggregating task traces to create task models that developer tools can use to assist end users in scripting personalized service robot tasks. Our dataset demonstrates that the interface is scalable to a large number of tasks and is easy for demonstrators to use. 

\begin{acks}
This work was supported by the National Science Foundation (NSF) award 1925043 and a University of Wisconsin–Madison Letters \& Science Community of Graduate Research Scholars Fellowship. This work was conducted while David Porfirio was affiliated with the University of Wisconsin–Madison and finalized while an NRC Postdoctoral Research Associate at the Naval Research Laboratory. 
\end{acks}

\bibliographystyle{ACM-Reference-Format}
\balance
\bibliography{biblio}

\end{document}